\documentclass[12pt]{article}
\usepackage[pdftex]{graphicx}
\usepackage{authblk}

\title{Usefulness of Higher-Order System-Size Correction for  Macromolecule Diffusion Coefficients: A Molecular Dynamics Study}

\begin{document}
\author[1]{Tomoya Iwashita}

\affil[1]{Department of Chemistry,
            Kyushu University, 
            Fukuoka,
            812-0395, 
            Japan}

\author[1]{Masaaki Nagao}
\author[2]{Akira Yoshimori}

\affil[2]{Department of Physics,
            Niigata University, 
            Niigata,
            950-2181, 
            Japan}
\author[3]{Masahide Terazima}
\affil[3]{Department of Chemistry,
            Kyoto University, 
            Kyoto,
            606-8502, 
            Japan}

\author[1]{Ryo Akiyama \thanks{Corresponding author. \it Email address\rm:\ rakiyama@chem.kyushu-univ.jp}}

\maketitle
\begin{abstract}
Yeh and Hummer's simplified estimation method has often been adopted to obtain diffusion coefficients for solute molecules using molecular dynamic simulation. However, the simplified formula is not necessarily valid when a small basic cell is used. Therefore, we conducted molecular dynamics simulations of aqueous protein solution to estimate the diffusion coefficient for the infinite dilution limit. We confirmed a deviation from the simplified formula in the simulation data and rationalized the discrepancy based on the unsimplified formula.
\end{abstract}

\newpage
\section{Introduction}
Einstein was interested in Brownian motion and proposed the Stokes--Einstein law by combining the fluctuation--dissipation theorem with fluid dynamics.\cite{Einstein1,Einstein2} The relation is shown as:
\begin{equation}
D = \frac{ k_{\rm B} \it T }{6\pi \eta R},
\end{equation}
where $D$, $k_{\rm B}$, $T$, $\eta$, and $R$ are the diffusion coefficient, Boltzmann constant, temperature, viscosity, and hydrodynamic radius for the Brownian particle, respectively. Although this relation was proposed over 100 years ago, it is still in use today to determine the molecular size in a fluid. However, it should be noted that the molecular size determined by the relation is the hydrodynamic radius $R$, which is not always the same as the radius of the particle itself. \par

Terazima et al. found that the difference can be highly variable using the transient grating method.\cite{Terazima1} They showed that the diffusion coefficients changed significantly upon conformation changes associated with reactions, even when the shape and volume changes of the proteins were minor.\cite{Terazima2, Terazima3, Terazima4, Terazima5, Terazima6, Terazima7, Terazima8} In the case of LOV2-linker, for example, the diffusion coefficient changes significantly upon photochemical reaction.\cite{Terazima2, Terazima3, Terazima4} If the observed changes in the hydrodynamic radii were assumed to be proportional to the radii of the proteins, the volume increase was calculated to be more than two times; however, such large volume changes were not found experimentally. The authors proposed that the hydrodynamic radius change is caused by the hydration change.\par

The change of diffusion coefficients due to the solvation change was studied theoretically by Nakamura et al.\cite{Yoshimori1, Yoshimori2, Yoshimori3, Yoshimori4, Yoshimori5}. The diffusion coefficients of a hard sphere in a binary hard sphere mixture were calculated and the fraction dependences were studied. These results indicate that the diffusion coefficient depends strongly on the solvation. On the other hand, we have not found a theoretical study for the correlation between the change in hydration and the drastic change in the diffusion coefficient for proteins. Molecular dynamics (MD) simulations seem to be suitable for the study of these changes; however, the traditional method of calculation may not be valid for estimating the diffusion coefficient.\par

In MD simulations, diffusion coefficients are highly dependent on the size of the simulation cell. Therefore, we need correction methods to estimate the diffusion coefficient for an infinite system $D_0$ in which this dependence vanishes. Correction methods have already been proposed\cite{YHM,FM}. In particular, Yeh and Hummer showed that the diffusion coefficient for the finite system under periodic boundary conditions $D_{\rm pbc}$ can be expressed by explicitly using physical quantities, in the following equation: 
\begin{equation}
\label{YH_n}
  D_{\rm pbc}=D_{0}^{\rm YH1} - \frac{k_{\rm B}\it T\xi}{6\pi\eta_{\rm sol} L},
\end{equation}
where $D_{0}^{\rm YH1}$ is the diffusion coefficient for the infinite system in the equation, $L$ is the side length of the cubic basic cell, $\xi$ is a constant equal to 2.837297 in the cubic basic cell, and $\eta_{\rm sol}$ is the shear viscosity of the solvent. 

Eq. (\ref{YH_n}) gives us adequate $D_0$ for a Lennard--Jones liquid even when the system has only 27 particles\cite{YHM}. This equation has been used in various systems to estimate the diffusion coefficients $D_0$\cite{Edward, Othonas, Tavares, Hummer1,Tokunaga1,Tokunaga2}. On the other hand, an additional correction term was shown as follows:
\begin{equation}
\label{YH_c}
  D_{\rm pbc}=D_0^{\rm YH2} - \frac{k_{\rm B}\it T\xi}{6\pi\eta_{\rm sol} L}+\frac{2 k_{\rm B}\it T R^{\rm2}}{9\eta_{\rm sol} L^3}.
\end{equation}
Yeh and Hummer introduced the additional correction term, namely, the third term in Eq. (\ref{YH_c}), based on Hasimoto's solution for the Stokes equation under periodic boundary conditions\cite{YHM_H}. $D_{0}^{\rm YH2}$ is the estimated diffusion coefficient for the infinite system. \par

The third term of Eq. (\ref{YH_c}) includes the hydrodynamic radius $R$. Unfortunately, the coefficient $D_0$ is needed to obtain the radius $R$. Therefore, the estimation of $D_0$ becomes difficult. Yeh and Hummer noted that the third term of Eq. (\ref{YH_c}) was relatively small for $R>L/2$\cite{YHM}. After the work, several studies were conducted to estimate diffusion coefficients using Yeh--Hummer's correction method with MD simulations. In most estimations, Eq. (\ref{YH_n}), which neglects the third term, has been adopted. The contribution of the third term was not crucial in these discussions on each phenomenon.

Notably, Eq. (\ref{YH_c}) can be a more suitable equation if accurate determinations of protein diffusion coefficients are desired. We can estimate the deviation given by the third term if we assume the values $R$ and $L$. The Eq. (\ref{YH_c}) can be rewritten in the form:
\begin{equation}
  D_{\rm pbc}=D_0 - \frac{k_{\rm B}\it T\xi}{6\pi\eta_{\rm sol} L}(1-\frac{4\pi R^2}{3\xi L^2}),
\end{equation}
where the percentage of the third term in the correction term can be expressed as $4\pi R^2/3\xi L^2$. Assuming that $D_0$ obeys the Stokes--Einstein law (Eq. (1)), Eq. (\ref{YH_c}) also can be expressed as follows:
\begin{equation}
  D_{\rm pbc}=D_0 \{1- \frac{R}{L}(\xi-\frac{4\pi R^2}{3L^2})\},
\end{equation}
where the percentage of the correction term in $D_0$ is expressed as $R/L(\xi-4\pi R^2/3L^2)$. Hence, the percentage of the third term in $D_0$ is the product of $4\pi R^2/3\xi L^2$ and $R/L(\xi-4\pi R^2/3L^2)$. \par

For the purpose of hydration change studies, we want the percentage contribution of the third term, namely $4\pi R^2/3\xi L^2 \times R/L(\xi-4\pi R^2/3L^2)$, to be less than 1 \% of $D_0$. When this condition is satisfied, the basic cell size $L$ must be larger than 7.4$R$. In the case of large solute molecules, conducting a MD simulation that satisfies the condition becomes very computationally expensive. For example, the radius of the LOV2 domain is about 3 nm\cite{Hisatomi}. If the hydrodynamic radius is the same (3 nm), we must prepare a basic cell for which $L$ is at least 22 nm. If we assume that the basic cell has one LOV2 domain, the number of solvent molecules is expected to be more than 3$\times10^5$. The calculation cost for $D_0$ becomes more expensive as the size of solute molecule increases. Then, Eq. (\ref{YH_c}) is useful in the present study.\par

In the present study, we will prepare a method to obtain an infinite dilution diffusion coefficient for a macromolecule $D_0$, using the data obtained from MD simulations for some small systems. Notably, the ratio of the number of protein molecules to the number of solvent molecules in the small systems is much larger than that in the infinitely dilute system. In other words, the concentration of the simulated systems is substantially higher than that for the infinitely dilute system, namely 0. However, Eq. (\ref{YH_c}) gives us a new strategy for the determination of $D_0$. First, we will examine the validity of Eq. (\ref{YH_c}) in a solution containing only one protein molecule, and the deviation from the simplified correction will become clear. This large deviation has not been shown previously. Next, we will introduce a small scheme for the estimation method based on Eq. (\ref{YH_c}) without the hydrodynamic radius $R$ and conduct numerical tests. Eq. (\ref{YH_c}) was used in a previous study\cite{Hummer2}, but $R$ was obtained by using other methods in that study. Differences between the present and previous studies will be highlighted. The current approach reduces the numerical cost from ``enormous'' to ``acceptable.''. Finally, the physical meaning of deviation will be discussed. \par

\newpage 
\section{Method}
It is difficult to obtain the diffusion coefficient $D_0$ for a molecule in an infinite system based on MD simulations because of the basic cell-size dependence. Several researchers have proposed methods that address this issue. The Fushiki method\cite{FM} is based on the linear response theory and linearized hydrodynamics. Fushiki showed that $D_{\rm pbc}$ can be expressed by the following equation:
 \begin{equation}
 \label{FM_e}
  D_{\rm pbc}=D_0^{\rm F} - \alpha\frac{1}{L},
 \end{equation}
where $D_0^{\rm F}$ is the diffusion coefficient for the infinite system, and $\alpha$ is a constant that is independent of the system size. Although MD simulations for the infinite system are not possible, we can estimate $D_0^{\rm F}$ by using this linear function of $1/L$ and $D_{\rm pbc}$ data calculated for various system sizes. $D_0^{\rm F}$ is given by the vertical intercept of the plot $D_{\rm pbc}$ vs. $1/L$.\par

A second approach is the Yeh--Hummer method.\cite{YHM} Fushiki mentioned $\alpha$ and another method of estimating the diffusion coefficient in his paper. Yeh and Hummer explicitly showed that \(\alpha=k_{\rm B}T\xi/ \rm 6 \pi\it\eta_{\rm sol}\) based on the Kirkwood--Riseman theory of polymer diffusion. Here, we obtain Eq. (\ref{YH_n}). Then, we obtained $D_0$ from $D_{\rm pbc}$ and $\eta_{\rm sol}$ using Eq. (\ref{YH_n}). Previous studies have reported that Eq. (\ref{YH_n}) is reasonable in various systems and basic cell sizes\cite{YHM, FM, Edward, Othonas, Tavares, Hummer1,Tokunaga1,Tokunaga2}. In principle, however, given the small size of the system, the correction value $D_0^{\rm YH1}$ must depend on $1/L$ due to the additional correction term (Eq. (\ref{YH_c})).\par

The Fushiki method and the simplified Yeh--Hummer method (Eq.(\ref{YH_n})) are based on the same correction principle. The results obtained from both should be consistent within a numerical error. We can thus use the Fushiki method to confirm the validity of the simplified Yeh--Hummer method.\par
 
\paragraph{MD simulations}
We performed the MD simulations under $NVT$ conditions to obtain $D_{\rm pbc}$ of a protein in aqueous solution. Two proteins, chignolin\cite{CGL} and $\beta$-lactoglobulin(BLG)\cite{BLG}, were examined. To satisfy the charge neutrality, Na$^+$ ions were dissolved into the solution. TIP4P2005 was adopted as a model of the water molecule.\cite{TIP4P2005} The all-atom optimized potentials for liquid simulations (OPLS-AA) force field was used for proteins and ions.\cite{OPLSAA} Ambient conditions, $T=298.15$ K and $P=1$ bar, were assumed. The temperature was maintained by using the modified Berendsen thermostat.\cite{V_rescale} The relaxation time was 0.1 ps. Each $L$ was determined by the average cell size of the $NPT$ ensemble ($T=298.15$ K, $P=1$ bar). The pressure was maintained by the Parrinello--Rahman barostat.\cite{barostat} The parameters are summarized in Table \ref{sys_para}. We also performed MD simulations of pure water under $NVT$ conditions to obtain $\eta_{\rm sol}$. All $NVT$ simulations were conducted for a total of 200 ns after equilibration. For all the MD simulations, we used GROMACS software with particle mesh Ewald summation.\cite{GROMACS,GROMACS2,PME} The equations of motion were integrated using the leap-frog algorithm with a time step of 2 fs.\par

 \begin{table}[ht]
 \centering
\caption{System parameters}
\label{sys_para}
\scalebox{1.0}{
\begin{tabular}[b]{c c c} 
\noalign{\global\arrayrulewidth=1pt}\hline
\multicolumn{1}{c}{\small the simulation}&\multicolumn{1}{c}{\small the number}&\multicolumn{1}{c}{\small the number} \\
\multicolumn{1}{c}{\small cell size}&\multicolumn{1}{c}{\small of}&\multicolumn{1}{c}{\small of} \\
\multicolumn{1}{c}{$L$[nm]}&\multicolumn{1}{c}{\small water molecules}&\multicolumn{1}{c}{\small $\rm{Na}^+$ ions} \\
\hline\noalign{\global\arrayrulewidth=0.4pt}
 &chignolin\textsuperscript{\emph{a}} & \\ \hline
  2.49092 & 471   & 2 \\
  2.94867 & 807   & 2 \\
  4.06195 & 2171  & 2 \\
  5.00068 & 4084  & 2 \\
  6.00696 & 7106  & 2 \\
  6.98039 & 11173 & 2 \\ 
  8.00418 & 16865 & 2 \\
  9.01075 & 24078 & 2 \\ \hline
 &$\beta$-lactoglobulin\textsuperscript{\emph{a}} & \\ \hline
  5.99188 & 6334   & 8 \\
  6.96927 & 10407  & 8 \\
  9.01441 & 23394  & 8 \\
 11.00137 & 43140  & 8 \\
 12.01388 & 56409  & 8 \\
 12.95841 & 70970  & 8 \\
 14.00551 & 89805  & 8 \\
 14.96803 & 109777 & 8 \\ \hline
 &pure water& \\ \hline
  2.48582 & 512  & 0 \\ \hline 
\end{tabular}
}
 \scalebox{1.0} {
  \textsuperscript{\emph{a}} The number of proteins is fixed to 1. ;
    }
\end{table}

\newpage
Each mean-square displacement $\langle|r(t)-r(0)|^2\rangle$ (MSD) of the proteins was calculated using the MD trajectory. $D_{\rm pbc}$ is shown as follows:
\begin{equation}
  D_{\rm pbc} = \frac{1}{6}\lim_{t \to \infty}\frac{\partial}{\partial t}\langle|r(t)-r(0)|^2\rangle,
\end{equation}
where $r(t)$ represents the position of the protein at time $t$. Each $D_{\rm pbc}$ was obtained from the slope of MSD. Each part from 40 ps to 80 ps (chignolin) and 60 ps to 100 ps (BLG) was used to obtain $D_{\rm pbc}$. 
Each shear viscosity $\eta$ was calculated using the Green--Kubo relation:\cite{viscosity1,viscosity2}
\begin{equation}
  \eta=\lim_{t\to \infty}\frac{V}{k_{\rm B}\it T}\int_{0}^t \frac{1}{5} \sum_i^5 \langle P_i(t)P_i(0) \rangle,
\end{equation}
where $V$ is the volume of the system and $P_i$ represents each of the five independent components of the traceless stress tensor, $P_{i=1\sim 5}$=\([(P_{xx}-P_{yy})/2, (P_{yy}-P_{zz})/2, P_{xy}, P_{yz}, P_{zx}]\). In the present study, each viscosity was estimated as the integral over 10ps. Diffusion coefficient and shear viscosity values were obtained as the average of over 1000 trials. The error bars in the figures were obtained from the standard error.\par

\section{Results and Discussion}
  \begin{figure}[ht]
  \centering
 \includegraphics[width=0.7\linewidth]{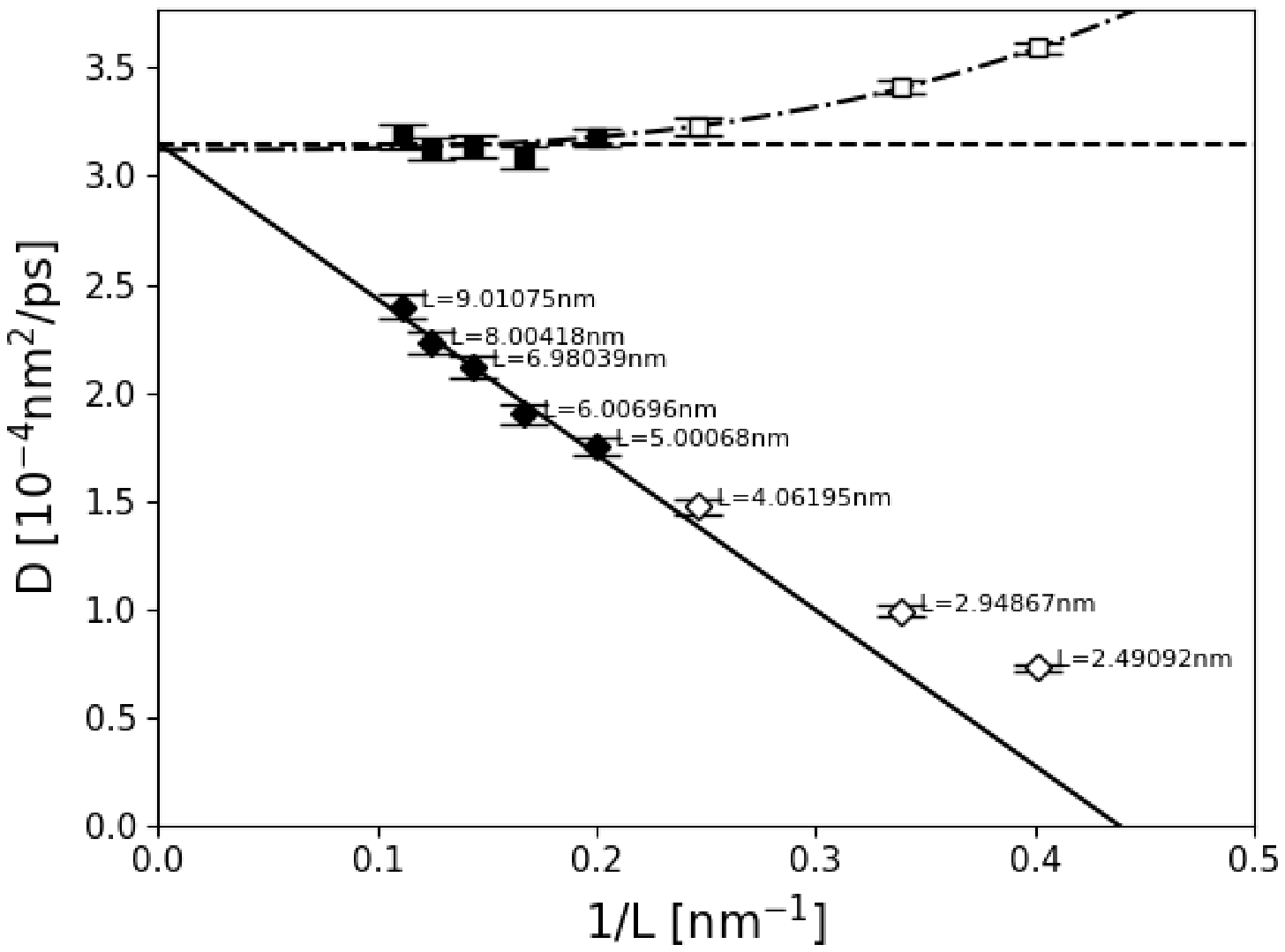}
 \caption{The diffusion coefficients of chignolin as a function of the inverse box length $1/L$ [$\rm{nm^{-1}}$] for $N=471\sim24078$ water molecules, sodium ions, and chignolin ($L=2.49092\sim9.01075$nm). Diamonds indicate the average of uncorrected diffusion coefficients $D_{\rm pbc}$ given by MD simulations for each box size. Squares indicate corrected diffusion coefficients in Eq. (\ref{YH_n}). The straight solid and dashed lines were obtained from the closed diamonds and closed squares, respectively. The dash--dot curve is plotted using Eq. (\ref{YH_fit}), where the parameters are determined using three open squares.}
 \end{figure}
 
  \begin{figure}[ht]
  \centering
 \includegraphics[width=0.7\linewidth]{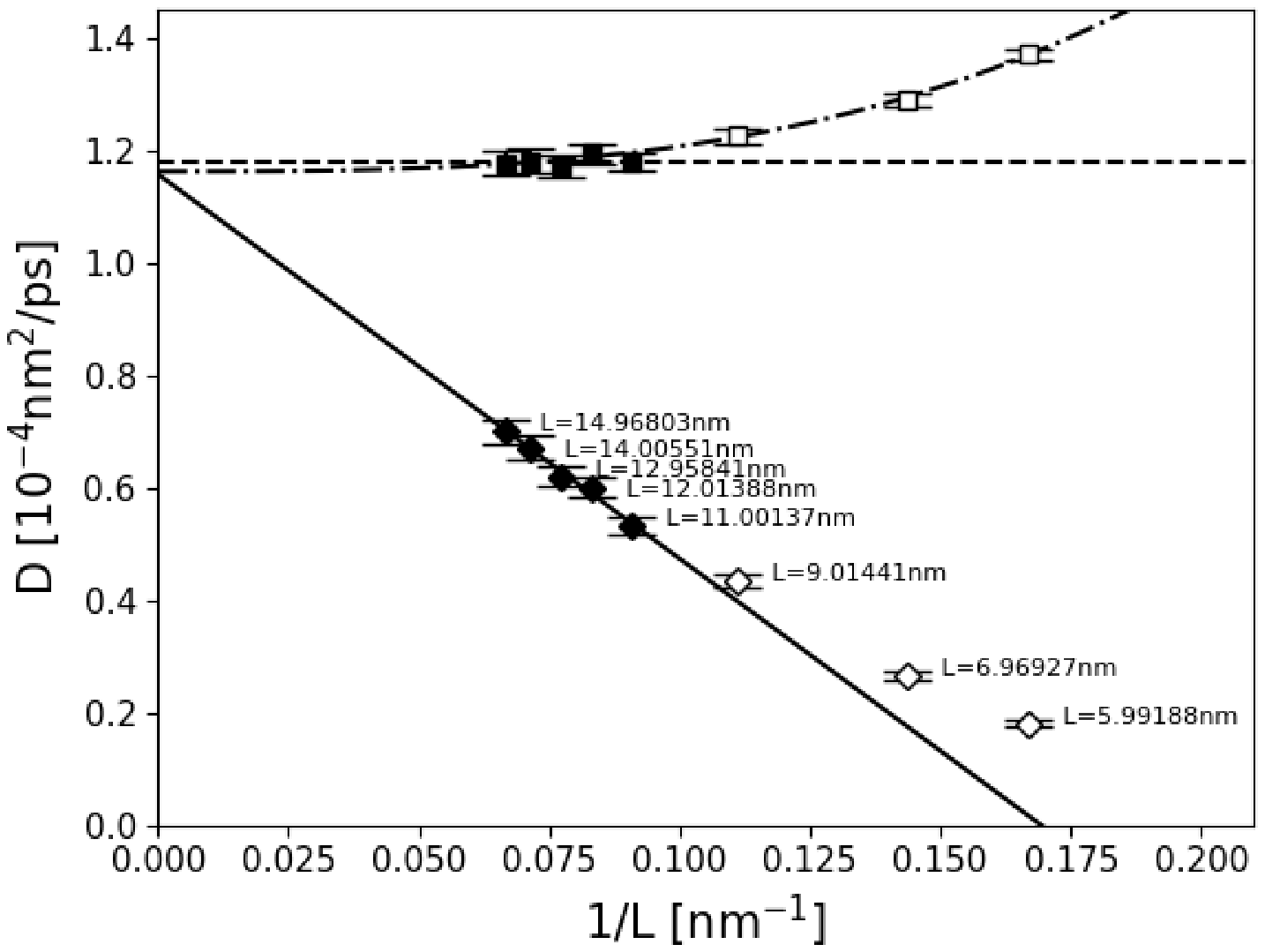}
 \caption{The diffusion coefficient of $\beta-$lactoglobulin as a function of the inverse box length $1/L$ [$\rm{nm^{-1}}$] for $N=6334\sim109777$ water molecules, sodium ions, and chignolin ($L=5.99188\sim14.96803$nm). Diamonds indicate the average of uncorrected diffusion coefficients $D_{\rm pbc}$ given by MD simulations for each box size. Squares are corrected diffusion coefficients in Eq. (\ref{YH_n}). The straight solid and dashed lines are obtained from the closed diamonds and closed squares, respectively. The dash--dot curve is plotted using Eq. (\ref{YH_fit}), where the parameters are determined using three open squares.}
 \end{figure}
 
Fig. 1 shows the system-size dependence of the diffusion coefficients $D_{\rm pbc}$ of chignolin in an aqueous solution (see closed and open diamonds). Squares (closed and open symbols) indicate corrected diffusion coefficients given by the simplified Yeh--Hummer method (Eq. (\ref{YH_n})) for each box size. We performed MD simulations to obtain the viscosity $\eta_{\rm sol}$ for water (8.68±0.06×$10^{-4}$pa$\cdot$s). The error bars for the diffusion coefficients are estimated from those for $D_{\rm pbc}$ and the viscosity $\eta_{\rm sol}$ for water. Given the small error bars, we think that the number of samples is sufficient; however, the plotted coefficients are not constant. As $1/L$ becomes larger, the deviation from the constant increases. The deviation becomes more pronounced in the region larger than 0.25 nm$^{-1}$. Fig. 2 shows the system-size dependence of the diffusion coefficients of a $\beta-$lactoglobulin (BLG) in an aqueous solution. A similar argument can be derived, with the deviation becoming more apparent in the region larger than 0.1 nm$^{-1}$.\par
 
If we adopt the simplified estimation using Eq. (\ref{YH_n}) with the $D_{\rm pbc}$ for the data indicated by open squares, the estimated $D_0$ deviates from the true value. This is confirmed by the results of calculations for $L>>R$ (see the closed squares). The simplified correction based on Eq. (\ref{YH_n}) with the $D_{\rm pbc}$ indicated by open squares overestimates the values. In contrast, the Fushiki method (Eq. (\ref{FM_e})) with the $D_{\rm pbc}$ indicated by open squares underestimates the values. These results indicate that a suitable solution for adequate estimation is obtained using large $L$ if we use Eqs. (\ref{FM_e}) or (\ref{YH_n}). This direct approach can be conducted if we want to obtain the diffusion coefficient of a small protein, such as chignolin or BLG. However, the application of the direct solution becomes harder as the size of the macromolecule increases.\par

Before discussing this issue further, we assess the more accurate expression given by Yeh--Hummer (Eq. (\ref{YH_c})). However, Eq.(\ref{YH_c}) contains the hydrodynamic radius $R$ itself. Here, we will obtain $R$ independently of Eq. (\ref{YH_c}). As mentioned above, we can prepare systems with large $L$ values in the case of a small protein, such as chignolin or BLG. For both proteins, the left-most set of five data points (closed squares) is constant. The average values $D^{\rm YH1}_{0}$ from each set of closed diamonds are shown in Table 2. We also applied the Fushiki method to the left five data points, namely, closed diamonds, and the estimated values $D^{\rm F}_{0}$ are also shown in Table 2. The values ($D^{\rm YH1}_{0}$ and $D^{\rm F}_{0}$) for each protein are almost the same. Here, $R$ values are obtained by using the Stokes--Einstein relation with each $D^{\rm YH1}_{0}$. The values $D_0$ were examined using Eq. (\ref{YH_c}) with the hydrodynamic radius $R$. The results are plotted as squares in Fig. 3(a),(b). Both plots are constant and consistent with the values $D^{\rm F}_0$ (dash--dot line) and $D^{\rm YH1}_{0}$ (dashed line). This agreement suggests that the deviations estimated by Eq.(\ref{YH_n}) from true $D_{0}$ can be corrected by Eq. (\ref{YH_c}). \par 
 
In the assessment of Eq. (\ref{YH_c}), we need a precise hydrodynamic radius $R$. Each protein has an averaged gyration radius $\langle R_g\rangle$ that can be easily obtained using MD simulations. Here, we cannot replace the hydrodynamic radius $R$ with the gyration radius $\langle R_g\rangle$. We obtain the values $\langle R_g\rangle$ (chignolin: 0.562 nm, BLG: 1.50 nm) and these gyration radii are used in the assessment of Eq.(\ref{YH_c}). The results are shown as crosses in Fig. 3(a),(b). The dependence of the deviation from the true values on system size remains. These results indicate that the gyration radii are not suitable for the assessment of Eq. (\ref{YH_c}). Experiments show that similar-sized proteins can have very different diffusion coefficients\cite{Terazima2, Terazima3, Terazima4, Terazima5, Terazima6, Terazima7, Terazima8}. This indicates that $\langle R_g\rangle$ is not suitable for the estimation of $D_0$.
 
 \begin{figure}[ht]
 \centering
 \includegraphics[width=0.7\linewidth]{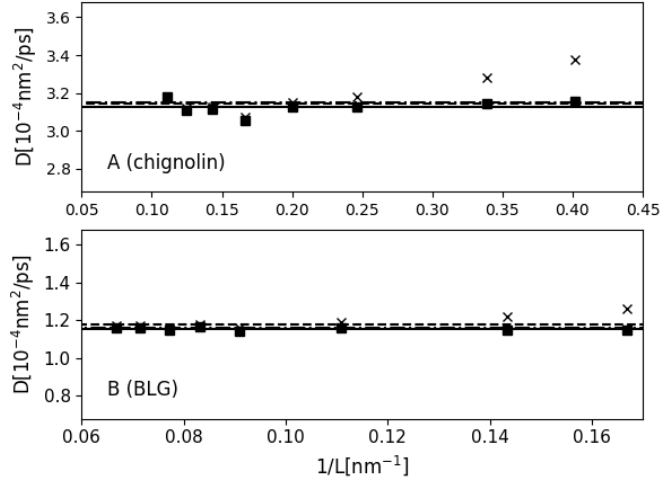}
 \caption{(a) The diffusion coefficient of chignolin as a function of $1/L$. (b) The diffusion coefficient of BLG as a function of $1/L$. In each figure, the squares are corrected diffusion coefficients estimated by using each uncorrected diffusion coefficient $D_{\rm pbc}$ and Eq. (\ref{YH_c}) with $R(D_{0}^{\rm YH2})$. The straight solid lines are obtained from the squares' average. The straight dash--dot and dashed lines show the previous results of the Fushiki and Yeh--Hummer methods, respectively. The crosses indicate corrected diffusion coefficients estimated using each uncorrected diffusion coefficient $D_{\rm pbc}$ and Eq.  (\ref{YH_c}) with the radius of gyration $\langle R_g\rangle$. }
\end{figure}

Next, we introduce a small scheme based on Eq. (\ref{YH_c}) without the hydrodynamic radius. We use Eq. (\ref{YH_n}) to remove $D_{\rm pbc}$ from Eq. (\ref{YH_c}). The above analysis showed that the estimated value in Eq.(\ref{YH_n}) depends on the system size. Then, the estimated value is replaced by $D_{\rm YH}(x)$, where $x=1/L^3$. We can remove $D_{\rm pbc}$ from Eq. (\ref{YH_c}) using the replaced equation, and obtain the following equation:
\begin{equation}
\label{YH_fit}
  D_{\rm YH}(x) = D_0 + \beta(R) x,
\end{equation}
where the coefficient $\beta (R)$ is $2k_{\rm B} T R^2 / (9\eta_{\rm sol})$. $D_0$ and $\beta (R)$ are obtained by fitting the data. Here, we name the estimated coefficient $D_0$ based on Eq. (\ref{YH_fit}) as $D^{\rm YH2}_0$. Although Eq. (\ref{YH_fit}) has two parameters, $D^{\rm YH2}_0$ and $\beta (R)$, it is essentially a one-parameter theory because there is a relation between $R$ and $D^{\rm YH2}_0$ (it is also essentially solving the cubic equation with one variable). However, we introduce one additional parameter, $\beta (R)$, to make it easier to fit the linear function $D^{\rm YH}(x)$. \par

The method based on Eq. (\ref{YH_fit}) is powerful because it gives us precise estimated values for $D_0$, which are very accurate even when the basic cells are small. Here, we estimated $D^{\rm YH2}_0$ and $\beta (R)$ using only data for small basic cells, that is, the data shown as open symbols. The dash--dot curves in Figs. 1 and 2 are plotted using the parameters. Although the closed symbols for large $L$ were not used in these parameter fittings, the curves are very accurate with all data.\par

Table \ref{ResultRD} includes $D_0$ and $R$ based on three equations, namely Eqs. (\ref{YH_n}), (\ref{FM_e}), and (\ref{YH_fit}). The agreements are good, which suggests two things. First, the estimation method based on the Eq. (\ref{YH_fit}) is powerful when we cannot prepare a basic cell that is large enough for other methods, such as the Fushiki (Eq. (\ref{FM_e})) and Yeh--Hummer (Eq. (\ref{YH_n})) methods. The agreement between $R(D^{\rm YH2}_0)$ and $R(\beta)$ suggests the consistency in Eq. (\ref{YH_fit}). However, the error of $R(D^{\rm YH2}_0)$ must be smaller than that of $R(\beta)$. The estimation method based on Eq. (\ref{YH_fit}) becomes more important as the macromolecule becomes larger.\par

\begin{table*}[ht] \centering

\caption{ Diffusion coefficients for $D_0$ and hydrodynamic radius $R$ in each method.\textsuperscript{\emph{a}} }
\label{ResultRD}
\scalebox{0.65}{
  \begin{tabular}{c c c c c c c c c c}\noalign{\global\arrayrulewidth=1pt}\hline
 \multicolumn{1}{c}{ }& \multicolumn{2}{c}{Fushiki Method}& \multicolumn{1}{c}{ }&\multicolumn{2}{c}{Yeh--Hummer Method(Eq. (\ref{YH_n}))}& \multicolumn{1}{c}{ }&\multicolumn{3}{c}{Eq. (\ref{YH_fit})(Eq. (\ref{YH_c}))}\\  \cline{2-3} \cline{5-6} \cline{8-10}
  \multicolumn{1}{c}{protein}&\multicolumn{1}{c}{$D_0^{\rm F}[\rm10^{{-}4}nm^{2}/ps]$}&\multicolumn{1}{c}{$R(D_0^{\rm F})$[nm]}& \multicolumn{1}{c}{}&\multicolumn{1}{c}{$D_0^{\rm YH1}[\rm10^{{-}4}nm^{2}/ps]$}&\multicolumn{1}{c}{$R(D_0^{\rm YH1})$[nm]}& \multicolumn{1}{c}{}&\multicolumn{1}{c}{$D_0^{\rm  YH2}[\rm10^{{-}4}nm^{2}/ps]$}&\multicolumn{1}{c}{$R(D_0^{\rm  YH2})$[nm]}&\multicolumn{1}{c}{$R(\beta)$[nm]}\\  \hline
  \multicolumn{1}{c}{chignolin}&\multicolumn{1}{c}{3.15}&\multicolumn{1}{c}{0.798}& \multicolumn{1}{c}{}&\multicolumn{1}{c}{3.14}&\multicolumn{1}{c}{0.800}& \multicolumn{1}{c}{}&\multicolumn{1}{c}{3.12}&\multicolumn{1}{c}{0.806}&\multicolumn{1}{c}{0.832}\\  
  \multicolumn{1}{c}{BLG}&\multicolumn{1}{c}{1.16}&\multicolumn{1}{c}{2.17}& \multicolumn{1}{c}{}&\multicolumn{1}{c}{1.18}&\multicolumn{1}{c}{2.13}& \multicolumn{1}{c}{}&\multicolumn{1}{c}{1.16}&\multicolumn{1}{c}{2.16}&\multicolumn{1}{c}{2.05}\\  
  \end{tabular}
}
\scalebox{0.75}{
  \textsuperscript{\emph{a}} $R(D_0)$ are obtained using the Stokes--Einstein law from each $D_0$.;
}
\end{table*}

The present study highlights the deviations from the simplified Yeh--Hummer formula. The physical picture based on the simulation data has not been discussed yet; however, we can discuss it based on the equations because the numerical   deviations are followed by the unsimplified Yeh--Hummer formula, namely Eq. (\ref{YH_c}).The agreement suggests that the deviation is caused by the system-size dependence of solvent viscosity. We will discuss this in the following paragraphs. We then discuss the third correction term in Eq.(\ref{YH_c}). Hashimoto proposed an equation as follows: \begin{equation}
  U = U_0 - \frac{B}{\mu \tau_0}
\end{equation}
where $\mu$, $\tau_0$, $U$, and $U_0$ are solvent viscosity equal to $\eta_{\rm sol}$, system volume equal to $L^3$, the mean flows under periodic boundary conditions in the finite basic cell and in the infinite basic cell, respectively\cite{YHM_H}. $B$ derived in Ref.24 depends on the hydrodynamic radius $R$, and $-B/ \eta_{\rm sol} L^3$ is the order of $4\pi R^3/(3L^3)$. The relation between $U$ and $U_0$ can be rewritten as:
\begin{equation}
\label{U_eq}
  U = U_0 ( 1 + \frac{4 \pi R^3}{3 L^3})
\end{equation}
If we substitute Eq. (\ref{U_eq}) into Stokes--Einstein law:
\begin{equation}
  D_{\rm YH} (x) = \frac{k_{\rm B} T}{F} U
\end{equation}
where $F$ is the Stokes's drag, we obtain the next relation:
\begin{equation}
\label{eq_YH_c}
  D_{\rm YH} (x) = k_{\rm B} T\frac{U_0}{F}  ( 1 + \frac{4 \pi R^3}{3 L^3})
\end{equation}
The approximation of Eq. (\ref{eq_YH_c}) is the same as that of Eq. (\ref{YH_fit}) (or Eq. (\ref{YH_c})). In this context, the third term indicates the dependence of mean flow on the system size. \par

 On the other hand, Reynolds number $Re$ also has the system-size dependence as follows,
\begin{equation}
 Re = \frac{\rho R U_0}{\eta_{\rm sol}} ( 1 + \frac{4 \pi R^3}{3 L^3})
\end{equation}
where $\rho$ is the solvent density. The hydrodynamic radius $R$ scales this equation. We can recognize that the system-size dependence of the mean flow is equivalent to that of solvent viscosity because of Reynolds’ law of similarity. Then, we can rewrite the size dependence of the solvent viscosity $\eta_{\rm pbc}$ as follows,
\begin{equation}
\label{i-ta_pbc}
 \eta_{\rm pbc}=\frac{\eta_{\rm sol}}{1+\frac{4 \pi R^3}{3 L^3}}
\end{equation}
If we substitute Eq. (\ref{i-ta_pbc}) into Stokes--Einstein law (Eq. (1)), we obtain Eq. (\ref{YH_fit}) itself.

 Moreover, Eq. (\ref{YH_c}) can be rewritten as follows,
\begin{equation}
\label{YH_c2}
  D_{\rm pbc}=D_{0} - \frac{k_{\rm B}\it T\xi}{6\pi\eta_{\rm int} L},
\end{equation}
where $\eta_{\rm int}$ is
\begin{equation}
 \eta_{\rm int}=\frac{\eta_{\rm sol}}{1-\frac{4 \pi R^2}{3 L^2 \xi}}.
\end{equation}
Comparing Eq. (\ref{YH_c2}) and Eq. (\ref{YH_n}), we can recognize $\eta_{\rm int}$ as the effective viscosity in the simplified Yeh--Hummer correction. Therefore, the correction in the Eq. (\ref{YH_c}) also means the system-size dependence of solvent effective viscosity $\eta_{\rm int}$ between the original and the replica proteins. Here, The two solvent viscosities $\eta_{\rm int}$ and $\eta_{\rm pbc}$ are defined in different ranges. $\eta_{\rm int}$ indicates the effective viscosity of the solvent in the region composed of all solvent molecules between the original and the replica proteins. In contrast, $\eta_{\rm pbc}$ indicates the viscosity of the bulk solvent in the region composed of the solvent molecules that is outside the hydrodynamic radius.\par

As discussed above, the unsimplified formula, Eq. (\ref{YH_c}), suggests the system-size dependence of the bulk solvent viscosity $\eta_{\rm pbc}$ and solvent effective viscosity $\eta_{\rm int}$. Unfortunately, we cannot directly extract the solution's viscosities from the simulation results. However, introducing a physical picture should enable verification based on the numerical calculation. Details will be reported in future articles based on this study because the calculation cost is expensive and an in-depth discussion of the results is beyond the scope of the present report.
 
 \begin{figure}[ht]
\centering
 \includegraphics[width=0.7\linewidth]{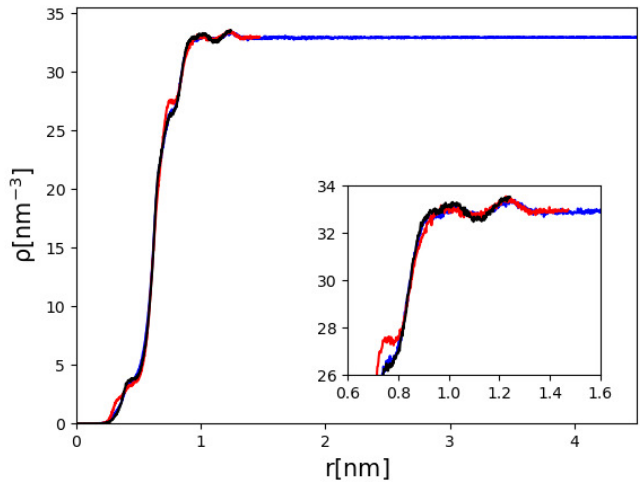}
 \caption{The density distribution function of water molecules (oxygen) around the center of the protein molecule (chignolin). The system in which $L$=2.49092 (black), 2.94867 (red), and 9.01075 nm (blue) is shown. In the inset, an enlargement of the region around $r=1.0$ nm is shown.}
 \end{figure}

 \begin{figure}[ht]
 \centering
 \includegraphics[width=0.7\linewidth]{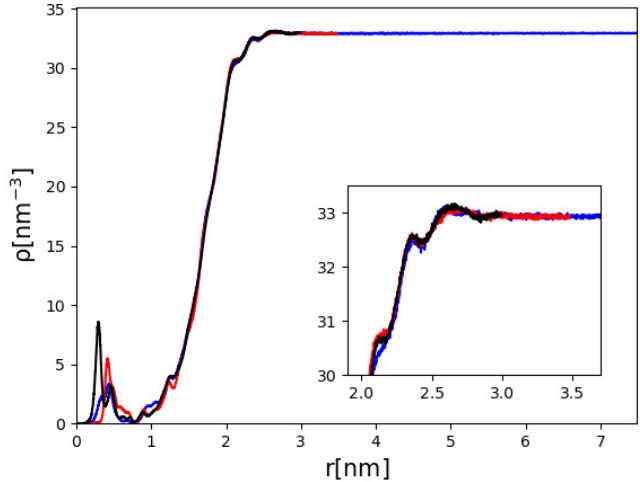}
 \caption{The density distribution function of water molecules (oxygen) around the center of the protein molecule (BLG). The system in which $L$=5.99188 (black), 6.96927 (red), and 14.96803 nm (blue) is shown. In the inset, an enlargement of the region around $r=2.0 $nm is shown.}
 \end{figure}
 
 \newpage
 Finally, we have discussed the hydrodynamic effect of the basic cell size. However, the finite size also affects the hydration structure. To discuss the effect of the static structure, we obtain the density distribution functions of water around the center of the protein. Fig. 4 shows the density distribution $\rho(r)$ around a chignolin. For $L= 2.94867$ and $9.01075$ nm, the densities approach that of the bulk value at the end of the function. By contrast, $\rho(r)$ does not approach the bulk value at the end when $L=2.49092$ nm. The hydration shell breaks at the peak of the oscillation around $r=1.2$ nm. Although the hydration shell is incomplete, the values for the diffusion coefficient obey the hydrodynamic law, Eq. (\ref{YH_c}). Fig. 5 shows the density distribution $\rho(r)$ around BLG. Because $\rho(r)$ oscillates around $r= 3.0$ nm, the smallest basic cell size , $L=5.99188$ nm, is not enough to construct the complete hydration shell in the box. However, the values of the diffusion coefficient obey Eq.  (\ref{YH_c}), again. These results suggest that the solvation structure effect is relatively minor when the macromolecule has some solvation layers in the simulation box.\par

\section{Conclusion}
The diffusion coefficient calculated by the MD simulation is strongly dependent on the basic cell size. Yeh and Hummer proposed a method to estimate the coefficient at the infinite cell-size limit.\cite{YHM} In this work, we carried out MD simulations. We examined the unsimplified (Eq. (\ref{YH_c})) and the simplified (Eq. (\ref{YH_n})) formulas to estimate the diffusion coefficients of solute molecules in the infinite system. In many studies, the simplified Yeh--Hummer formula (Eq. (\ref{YH_n})) has been adopted and generally gives us an adequate diffusion coefficient for the infinite system when the system size is large enough. However, the MD data also showed the diffusion coefficient corrected by the simplified formula depended on the basic cell size. The deviation was observed when the basic cell size was small. This deviation was systematic and followed by the unsimplified Yeh--Hummer formula, namely Eq. (\ref{YH_c}). We discussed the physical picture of this deviation based on the unsimplified formula. We showed that the deviation relates to the size dependence of solvent viscosity. A formula Eq. (\ref{YH_fit}), equivalent to Eq. (\ref{YH_c}), is more powerful in the estimation based on the simplified formula because this approach greatly reduces the calculation cost of the diffusion coefficients for solute molecules at infinite dilution. The reduction becomes more significant as the solute size increases. \par

\section{Acknowledgments}
We thank Drs Y. Uematsu of Kyushu Institute of Technology and Y. Nakamura of Niigata University for their comments. The computations were performed using the Research Center for Computational Science, Okazaki, Japan, and the Research Institute for Information Technology, Kyushu University, Japan, respectively. This work was supported by the Japan Society for the Promotion of Science KAKENHI (grant nos. JP21K18604, JP19H01863, JP19K03772, JP18H03673, JP18K03555, JP17H03008,  JP16H00774, JP16K05512, JP15K05249, and JP26102533).

 \bibliographystyle{unsrt} 
 \bibliography{ref}






\end{document}